\documentclass[conference]{IEEEtran}
\IEEEoverridecommandlockouts
\usepackage{cite}
\usepackage{amsmath,amssymb,amsfonts}
\usepackage{algorithmic}
\usepackage{graphicx}
\usepackage{textcomp}
\usepackage{xcolor}
\def\BibTeX{{\rm B\kern-.05em{\sc i\kern-.025em b}\kern-.08em
    T\kern-.1667em\lower.7ex\hbox{E}\kern-.125emX}}

\newcommand \ignore[1]{}
\newcommand*{\affaddr}[1]{#1} % No op here. Customize it for different styles.
\newcommand*{\affmark}[1][*]{\textsuperscript{#1}}
\newcommand*{\email}[1]{\texttt{#1}}
\usepackage{url}
\usepackage{hyperref}
\usepackage{colortbl}

\usepackage{caption}
\usepackage{subcaption}

\begin{document}

\title{
%\thanks{Thanks to Clifford B. Kinley Trust for funding this research.}
Audio-Based Classification of Insect Species Using Machine Learning Models: Cicada, Beetle, Termite, and Cricket
}

\author{Manas V Shetty\affmark[1], Yoga Disha Sendhil Kumar\affmark[2]\\
\affaddr{\affmark[1]Computer and Information Technology Department, \\
Purdue University, IN 47907}\\
\email{\{\affmark[1]shetty76,\affmark[2]ysendhil}\}@purdue.edu}

\maketitle

\begin{abstract}
This project addresses the challenge of classifying insect species—Cicada, Beetle, Termite, and Cricket—using sound recordings. Accurate species identification is crucial for ecological monitoring and pest management. We employ machine learning models such as XGBoost, Random Forest, and K-Nearest Neighbors (KNN) to analyze audio features, including Mel Frequency Cepstral Coefficients (MFCC). The potential novelty of this work lies in the combination of diverse audio features and machine learning models to tackle insect classification, specifically focusing on capturing subtle acoustic variations between species that have not been fully leveraged in previous research. The dataset is compiled from various open sources, and we anticipate achieving high classification accuracy, contributing to improved automated insect detection systems.
\end{abstract}

\begin{IEEEkeywords}
Insect classification, audio analysis, machine learning, XGBoost, Random Forest, K-Nearest Neighbors (KNN), Mel Frequency Cepstral Coefficients (MFCC), acoustic feature extraction, species identification
\end{IEEEkeywords}

\section{Motivation}

Early identification of harmful insect species in crops can prevent significant damage and mitigate crop loss, with automated systems enabling early detection and intervention to minimize the risk of widespread infestation. Tracking insect species diversity in ecosystems is essential for ongoing assessments without disturbing the environment. Insects like \textit{Barkbeetles}, \textit{Cicadas}, \textit{Crickets}, and \textit{Termites} play crucial roles in maintaining ecological balance, and their populations need to be monitored to understand broader environmental changes. Monitoring population shifts provides valuable insight into the impacts of climate change and pollution, with \textit{Cicada} populations serving as indicators of habitat loss and other ecological stresses. Real-time insect monitoring through automated acoustic tools enhances pest management by enabling timely interventions and reducing reliance on chemical treatments, ultimately minimizing damage to crops and structures.

Proactive pest management is essential to avoid the extensive costs associated with pest infestations, which can reach billions annually. For example, \textit{Termites} alone cause an estimated \$5 billion in damage in the U.S. each year \cite{developmentaid_pests_losses_2024}, while \textit{Barkbeetles} devastate forests and timber industries. Insect pests contribute to up to 40\% of global crop losses annually, resulting in over \$220 billion in economic losses due to pests and plant diseases \cite{developmentaid_pests}. In addition, the decline of insect species—estimated at 40\% globally—underscores the economic and ecological importance of biodiversity monitoring, with \textit{Cicadas} being key pollinators in agricultural ecosystems \cite{fao_news_story}. As \textit{Crickets} are sensitive to environmental stressors, monitoring their populations can offer valuable insights into the broader impacts of climate change and pollution \cite{fao_plant_health}. Early detection through automated systems can reduce these costs significantly, providing real-time alerts for pests like \textit{Termites}, which, if left unchecked, can lead to substantial long-term damage. 

By implementing automated monitoring systems for species like \textit{Cicadas}, \textit{Barkbeetles}, \textit{Crickets}, and \textit{Termites}, farmers and pest control agencies can effectively mitigate the economic and ecological toll of these pests. With the ability to detect up to 90\% of targeted insect activity, these systems help in timely interventions that prevent widespread damage to crops, forests, and infrastructure, ultimately supporting sustainable pest control strategies \cite{developmentaid_pests_2021}.

\section{Related Work}
While some researchers have been using image-based insect identification and some researchers have been using acoustic data for person identification \cite{dibbo2022phone, cheung2020continuous}, respiratory disease monitoring \cite{vhaduri2019towards, dibbo2021effect}, sleep health monitoring \cite{vhaduri2020nocturnal, vhaduri2018impact, chen2020estimating}, mental health and well-being management \cite{kim2020understanding, vhaduri2021deriving, vhaduri2021predicting}, insect identification from acoustic data did not receive much attention in the past. Recently, the classification of insect species using audio signals has gained increasing attention due to its potential in ecological monitoring and pest management. Previous studies have explored various methodologies for feature extraction, machine learning models, and dataset preparation to achieve accurate classification.

Acoustic signal analysis has been a core area of focus. Unique sounds produced by insects, such as chirps or stridulations, are distinct identifiers for species. Traditional approaches often use Mel Frequency Cepstral Coefficients (MFCCs) to represent these sounds. However, recent studies have highlighted the limitations of MFCCs for high-frequency insect signals, leading to the exploration of alternative techniques such as Linear Frequency Cepstral Coefficients (LFCCs) and adaptive waveform-based methods like LEAF. A notable example demonstrated the effectiveness of fusing MFCCs and LFCCs, achieving classification accuracy of up to 98.07\% for over 300 species \cite{acoustic_insect_mfcc_lfcc,adaptive_insect_classification}.

Deep learning has also contributed significantly to bioacoustics. Techniques such as Convolutional Neural Networks (CNNs) and EfficientNet have shown promise in extracting both spectral and temporal features. The Dual-Frequency and Spectral Fusion Module (DFSM), in particular, enhances accuracy by capturing intricate insect sound characteristics. These methods are vital for processing large datasets and overcoming challenges like noise and overlapping signals \cite{adaptive_insect_classification,dfsm_insect_classification}.

Traditional machine learning models, such as Random Forest and Support Vector Machines (SVM), continue to play an essential role in small-scale datasets or scenarios requiring explainable models. These models often rely on handcrafted features such as MFCCs and STFT-based spectrograms, which effectively capture key characteristics of insect sounds \cite{dfsm_insect_classification,ml_insect_bioacoustics}.

The availability of diverse and annotated datasets has been pivotal in advancing this field. Studies have emphasized the importance of high-quality datasets, such as the curated collections of insect sounds from orthopteran species and bioacoustic datasets like ESC-50. These datasets provide critical resources for training and validating models in real-world applications, ranging from biodiversity studies to agricultural pest control \cite{acoustic_insect_mfcc_lfcc,ml_insect_bioacoustics,framework_insect_sound_analysis}.

While significant progress has been made in the field of audio-based insect classification, many existing methods focus on either limited species or lack the integration of diverse feature extraction techniques and machine learning models. Our work aims to bridge these gaps by combining Mel Frequency Cepstral Coefficients (MFCCs), and advanced models such as XGBoost and Random Forest. By leveraging these techniques, our approach seeks to capture subtle acoustic variations and improve classification accuracy. This work has the potential to enhance automated insect monitoring systems, contributing to more efficient ecological monitoring, biodiversity assessment, and pest management strategies.

\section{Approach / Method}

\subsection{Problem Statement}
The primary challenge of this project is to accurately classify insect species—Cicada, Beetle, Termite, and Cricket—based on their unique audio signatures. Distinguishing between these species using sound recordings is critical for applications in ecological monitoring and pest management, where precise identification can inform conservation strategies and pest control measures.

\subsection{Proposed Models}
We will employ three machine learning models for classification: 

\begin{itemize}

    \item \textit{Random Forest}: This model is selected due to its robustness against overfitting and its ability to manage high-dimensional data effectively. Random Forest builds multiple decision trees and aggregates their predictions, enhancing model accuracy and stability in classifications ~\cite{datacamp2022machinelearning}.
    
    \item \textit{K-Nearest Neighbors (KNN)}: KNN is included for its simplicity and effectiveness in classification tasks. It classifies new data points based on the majority class of their nearest neighbors, making it a good choice for capturing local patterns in audio data ~\cite{datacamp2022machinelearning}.

    \item \textit{Decision Tree}: Decision Trees (DTs) are a versatile, non-parametric supervised learning method utilized for both classification and regression tasks. They operate by creating a model that predicts the value of a target variable based on simple, interpretable decision rules derived from the input features. Represented as a tree-like structure, DTs allow for piecewise constant approximations of complex functions, making them easy to visualize and understand\cite{scikit_learn_decision_trees}.

\end{itemize}

\subsection{Feature Sets}
The feature we will extract from the audio recordings include:

\begin{itemize}
    \item \textit{Mel Frequency Cepstral Coefficients (MFCC)}: MFCCs are widely used in audio processing and provide a representation of the short-term power spectrum of sound. They are particularly effective for capturing the timbral characteristics of audio signals, making them suitable for distinguishing between different insect sounds.
    
\end{itemize}

\section{Dataset Description}
\begin{figure}
    \centering
    \begin{subfigure}[b]{0.45\linewidth}
        \centering
        \includegraphics[width=\linewidth]{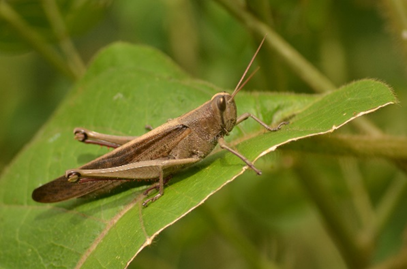}
        \caption{Cricket}
        \label{fig:cricket}
    \end{subfigure}
    \hfill
    \begin{subfigure}[b]{0.45\linewidth}
        \centering
        \includegraphics[width=\linewidth]{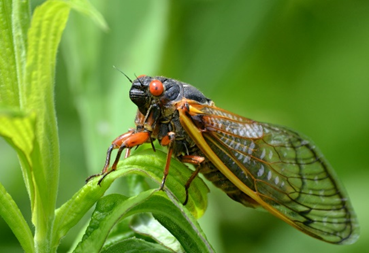}
        \caption{Cicada}
        \label{fig:cicada}
    \end{subfigure}
    \vfill
    \begin{subfigure}[b]{0.45\linewidth}
        \centering
        \includegraphics[width=\linewidth]{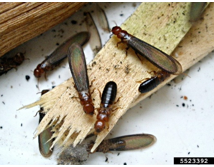}
        \caption{Termite}
        \label{fig:termite}
    \end{subfigure}
    \hfill
    \begin{subfigure}[b]{0.45\linewidth}
        \centering
        \includegraphics[width=\linewidth]{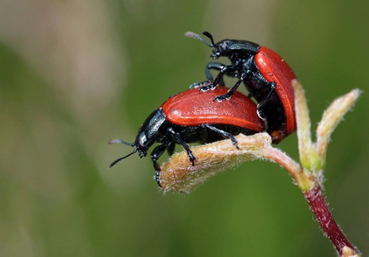}
        \caption{Bark Beetle}
        \label{fig:beetle}
    \end{subfigure}
    \caption{(a) Cricket, (b) Cicada, (c) Termite, and (d) Bark Beetle.}
    \label{fig:insects}
\end{figure}

Figure \ref{fig:insects} illustrates the different insect species included in the dataset.

\begin{itemize}
    \item \textit{Cricket}: The ESC-50 dataset contains 5-second-long audio recordings of environmental sounds, and for crickets, the dataset captures their distinct chirping or stridulation ~\cite{piczak2015esc}. Crickets generate these sounds by rubbing specialized structures on their wings together. This acoustic behavior is primarily linked to mating calls or territorial signals. The frequency and rhythm of cricket chirps can vary based on factors like species and temperature ~\cite{zhao2021connecting}.

    \item \textit{Cicada}:  The dataset contains 335 audio files across
 32 sound-producing insect species, totaling 57 minutes.
 It was designed for training neural networks to clas
sify insect species, comparing adaptive waveform-based
 frontends with mel-spectrograms for feature extraction.
 The dataset includes two subsets: 147 recordings from
 nine Orthoptera species and 188 recordings from 23
 Cicadidae species ~\cite{paperswithcode2022adaptive}. All files were resampled to 44.1 kHz
 mono WAV format and manually curated to remove noise
 interference. Annotations provide species information and
 training splits for neural network development.

    \item \textit{Termite}: The ARS Center for Medical, Agricultural, and Veterinary Entomology sound library includes a dataset featuring movement and feeding sounds of termites, specifically Coptotermes formosanus (Formosan termite) and Reticulitermes spp. (Eastern subterranean termite) ~\cite{mankin1998acoustic}. The sounds consist of head-banging and other movement-related noises produced by termites while foraging or defending their colonies. These audio files are selected from relatively noise-free sections; however, some recordings include typical background sounds mixed with insect activity to simulate real-world conditions. All recordings are provided in .wav format and have been resampled for consistency to support machine learning models and other research on insect detection and infestation control.

    \item \textit{Bark Beetle}: For the bark beetle dataset, it includes
 360 individualized acoustic signals from two species,
 Hylurgus ligniperda and Hylastes ater. Recordings were
 taken from 10 individuals (5 of each species) placed
 on the bark and within the phloem tissues of Pinus
 radiata logs. The beetles were recorded at nine different
 distances from the stridulating insect (5 to 100 cm). Each
 recording is 1-minute long and was captured using a
 352A24 monoaxial accelerometer and a 744T recorder at
 a sampling frequency of 44.1 kHz, 48 dB gain, and 24
bit resolution \cite{mankin1998acoustic}. This dataset is crucial for understanding
 the communication and behavior of bark beetles and can
 be used to train models for species detection based on
 acoustic signals
 \end{itemize}

\section{Data Preprocessing}
This section outlines the steps taken to prepare audio data for the classification of insect species, specifically Bark Beetle, Cricket, Cicada, and Termites. The preprocessing involved segmentation, and instantiation of the audio data to enhance its quality for model training. 

\subsection{Segmentation}
\begin{figure}
    \centering
    \includegraphics[width=0.45\textwidth]{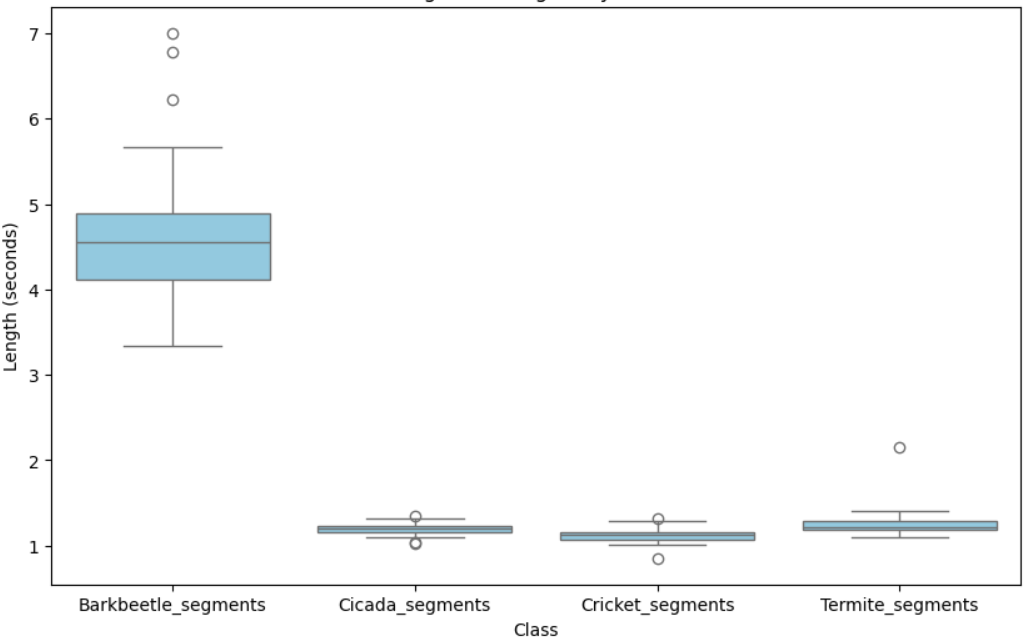}
    \caption{Segment Lengths by Class}
    \label{fig:boxplot}
\end{figure}

Segmentation involves dividing an audio recording into smaller, meaningful units. In this project, each audio file was segmented into multiple parts based on species-specific sounds. Using Audacity software, the recordings were divided by class, resulting in segments of varying lengths across classes. Each segment represents a meaningful audio unit, such as an insect’s characteristic sound, while excluding irrelevant sections like silence or background noise. Figure \ref{fig:boxplot} shows the segment lengths for each class.

\subsection{Instantiation}
\begin{figure}
    \centering
    \includegraphics[width=0.45\textwidth]{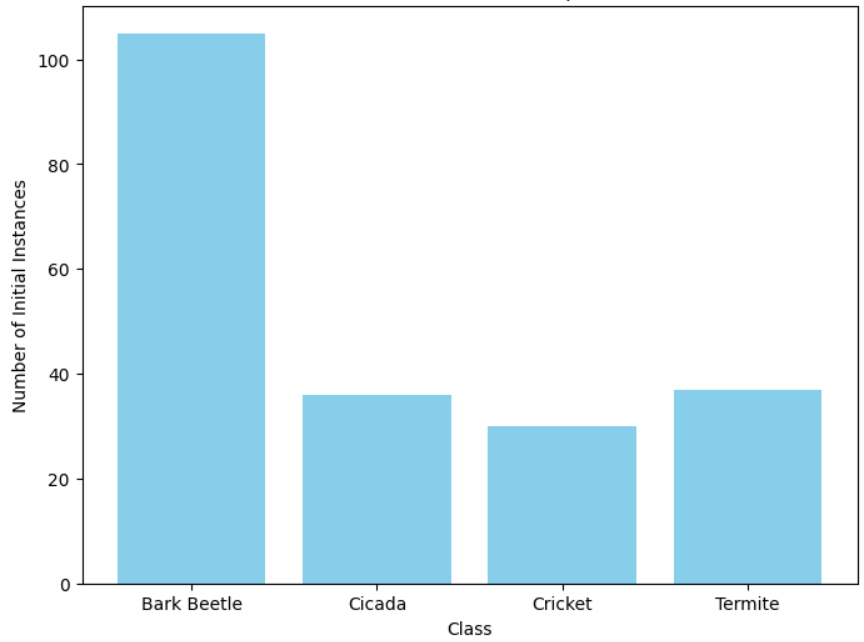}
    \caption{Number of Initial Instances Per Class}
    \label{fig:bar}
\end{figure}

After segmentation, each segment was treated as an individual instance. In this context, an instance refers to a data point or sample that will be used for training or testing a machine learning model. Instances were created from the segmented audio for each of the four classes—Bark Beetle, Cricket, Cicada, and Termites. All instances were standardized to have even lengths of 1 sec to ensure consistency in further processing steps and were labeled based on their respective classes. Figure \ref{fig:bar} shows the segment lengths for each class.

\section{Feature Extraction}

In this part of the project, 30 random instances were picked per class for feature extarction and modelling. Feature extraction was conducted to transform raw audio files into a structured format capturing essential acoustic properties, allowing for effective classification. Specifically, Mel Frequency Cepstral Coefficients (MFCCs) were extracted from each 1-second audio instance. MFCCs are widely used in audio processing and machine learning for their ability to represent spectral properties of sound, making them valuable for distinguishing between different audio classes.

The feature extraction process involved several steps. First, the base directory was defined along with the folders containing segmented audio instances for the four insect species—Bark Beetle, Cicada, Cricket, and Termite. Each folder contains preprocessed 1-second audio segments that serve as individual data instances. For each folder, the code iterates through a maximum of 30 audio files, loading each file while retaining the original sampling rate.

For each audio file, 40 MFCC features were computed, from which the mean and standard deviation of each MFCC coefficient were calculated across the time dimension. These statistics capture the central tendency and variability of the MFCCs within each audio segment, providing a compact yet informative representation of the sound profile. The extracted features, along with the class label, were then stored in a structured list and converted into a DataFrame for further analysis.

\begin{figure}
    \centering
    \includegraphics[width=0.45\textwidth]{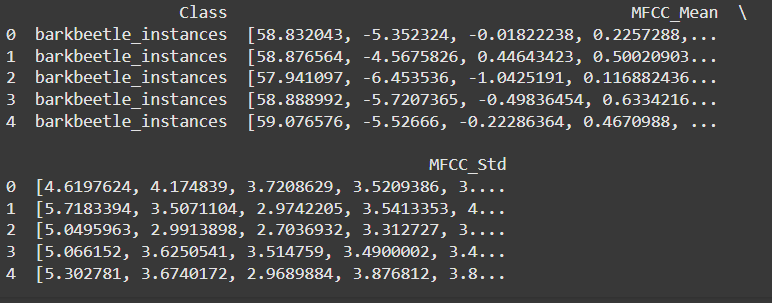}
    \caption{First Few Rows of the MFCC DataFrames}
    \label{fig:mfcc}
\end{figure}

\section{Modeling}

\subsubsection{Decision Tree}

A Decision Tree model was implemented to classify insect audio instances into four classes: Cicada, Cricket, Termite, and Bark Beetle. The Decision Tree algorithm is a non-parametric supervised learning method, ideal for tasks involving labeled data where the model must learn complex patterns for accurate classification.

The feature extraction process provided Mel Frequency Cepstral Coefficients (MFCC) as features, specifically the mean and standard deviation values of 40 MFCCs per audio instance. These MFCC values were flattened into individual columns and combined to form the feature set 
X. Each instance was labeled according to its class, which was then encoded numerically using a Label Encoder to create the target vector 
y.

The dataset was split into training and testing sets in an 80-20 ratio, ensuring that the model could be evaluated on previously unseen data. A Decision Tree Classifier with a random state of 42 was then initialized and trained on the training set. After training, predictions were made on the test set, and the model's accuracy, along with a classification report, was computed.

\begin{figure}
    \centering
    \includegraphics[width=0.45\textwidth]{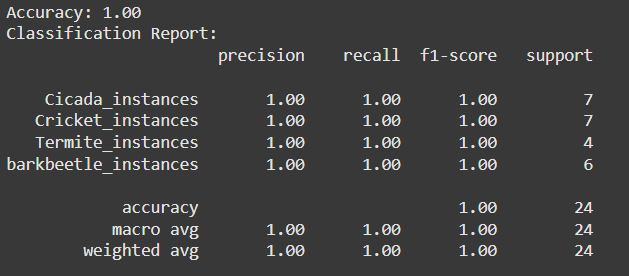}
    \caption{Decision Tree Results}
    \label{fig:dt}
\end{figure}

\begin{figure}
    \centering
    \includegraphics[width=0.45\textwidth]{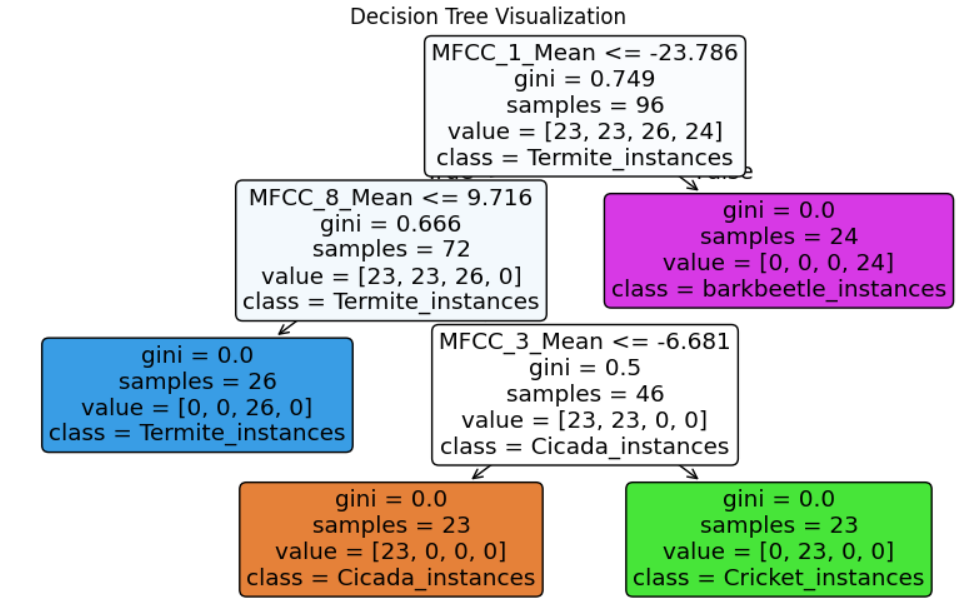}
    \caption{Decision Tree Visualization}
    \label{fig:dtv}
\end{figure}

As shown in the classification report (see Figure~\ref{fig:dt}
), the model achieved perfect scores with 1.00 precision, recall, and F1-score across all classes, indicating that it successfully distinguished between all insect types. Figure~\ref{fig:dtv}
 provides a visualization of the trained Decision Tree, highlighting the feature splits and class predictions.

\subsubsection{Random Forest}

A Random Forest classifier was implemented to classify insect audio instances into four classes: Cicada, Cricket, Termite, and Bark Beetle. The Random Forest algorithm, known for constructing multiple decision trees and combining their outputs, enhances the model's accuracy and robustness, making it particularly effective for complex datasets.

The same MFCC-based feature extraction approach used for the Decision Tree model was applied here. For each 1-second audio instance, the mean and standard deviation of 40 MFCC coefficients were calculated, forming the structured feature set X. The class labels were encoded using a Label Encoder to create the target vector y, and the dataset was split into training and testing sets in an 80-20 ratio.

\begin{figure}
    \centering
    \includegraphics[width=0.45\textwidth]{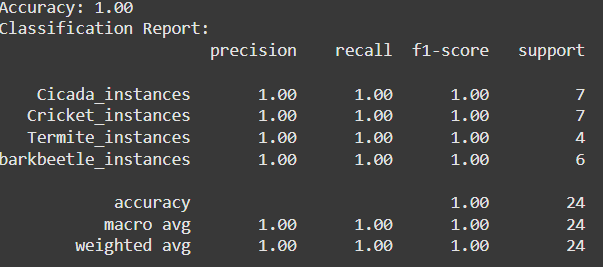}
    \caption{Random Forest Results}
    \label{fig:rf}
\end{figure}

Using 100 estimators and a random state of 42, the Random Forest model was trained and evaluated on the test set. The classification report (see Figure~\ref{fig:rf}) shows that the model achieved perfect accuracy with a precision, recall, and F1-score of 1.00 for all classes, indicating highly effective classification.

\begin{figure}
    \centering
    \includegraphics[width=0.45\textwidth]{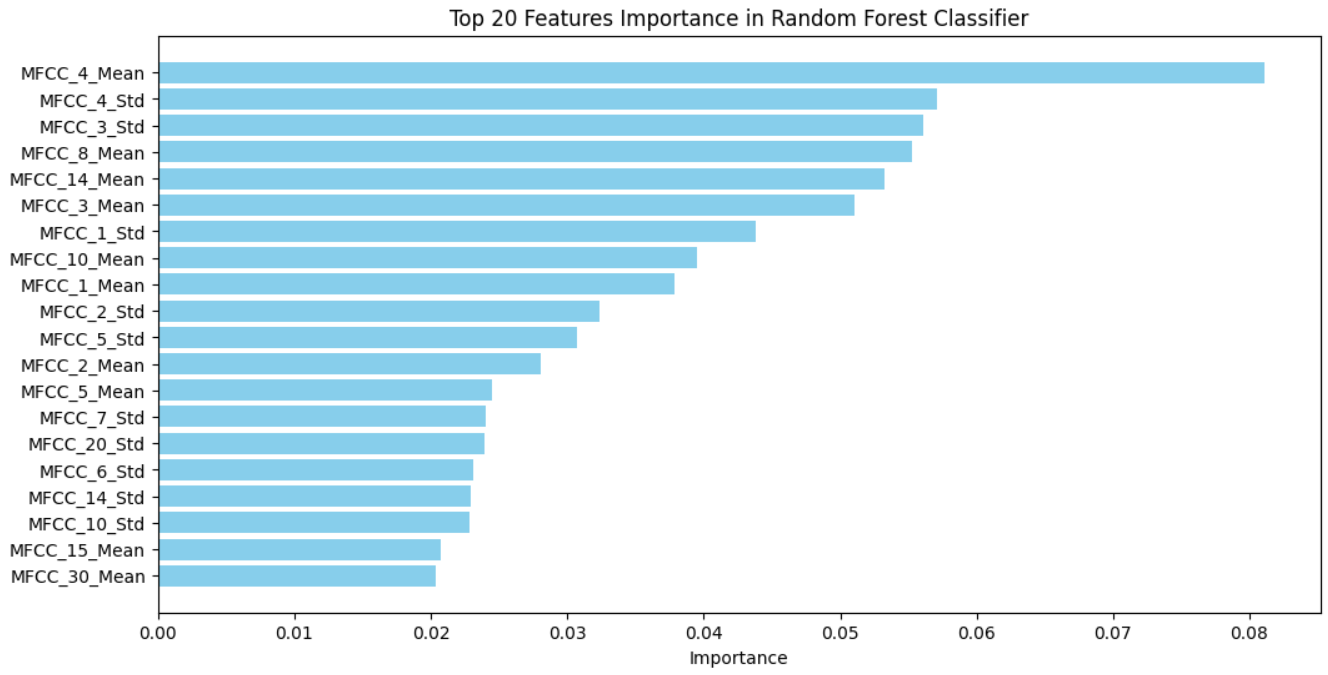}
    \caption{Top 20 Features Importance in Random Forest Classifier}
    \label{fig:rfv}
\end{figure}

Additionally, feature importance was analyzed to determine which MFCC features most significantly contributed to the model’s predictions. The top 20 important features, as identified by the Random Forest algorithm, are visualized in Figure~\ref{fig:rfv}, highlighting the MFCC features most relevant to insect sound classification.

\subsubsection{K-Nearest Neighbors (KNN)}

The K-Nearest Neighbors (KNN) classifier was employed as an alternative model for classifying insect audio instances. This algorithm classifies each test instance based on the majority class among its k-nearest neighbors in the feature space, where k is a predefined parameter. For this implementation, k was set to 5, providing a balanced trade-off between model complexity and generalization.

The MFCC features used for the Decision Tree and Random Forest models were also utilized for the KNN model, with each 1-second audio segment represented by the mean and standard deviation of 40 MFCC coefficients. These features formed the input feature set X, and the insect classes served as the target variable y. Following the standard 80-20 split for training and testing, the model was trained and evaluated on the test set.

\begin{figure}
    \centering
    \includegraphics[width=0.45\textwidth]{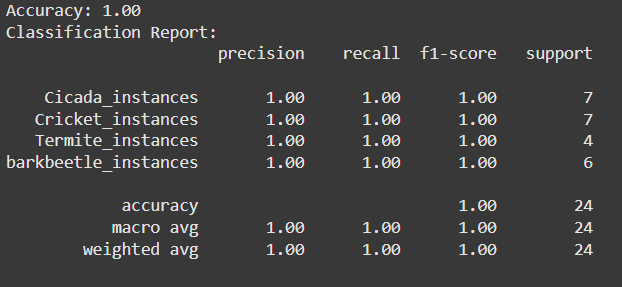}
    \caption{KNN Results}
    \label{fig:knn}
\end{figure}

The classification report (see Figure~\ref{fig:knn}) reveals that the KNN model achieved an impressive accuracy of 1.00, with a precision, recall, and F1-score of 1.00 for all classes. This high performance indicates that the KNN model was effective in accurately classifying each insect type based on the MFCC features.

\section{Updated Feature Extraction}

In the subsequent stages, we continued using Mel-frequency cepstral coefficients (MFCCs) for feature selection due to their effectiveness in distinguishing acoustic patterns. Our updated approach to feature selection involved refining the MFCC extraction process to optimize the input for our classification models. We experimented with extracting different numbers of MFCC coefficients—specifically, 10, 20, 30, and 40 coefficients—to determine the optimal number that maximizes classification accuracy. 

For each audio clip, we used the librosa library to compute the MFCC coefficients for every sample. To reduce dimensionality, we calculated the mean of each coefficient across the time axis, creating a fixed-length vector for each audio instance. This vector was then organized into a feature matrix, denoted as \( X \), and standardized for model training and testing. The refined feature vectors allowed us to capture unique acoustic characteristics of each insect species, providing a solid foundation for classification tasks and enabling us to evaluate the impact of varying the number of MFCC features on model performance.

\section{Updated Modeling with Decision Tree Model}

With feature selection complete, we advanced to the classification task, centering our approach on a Decision Tree model using 100 percentage of the instances for training and testing purposes. Decision Trees were chosen for their interpretability and ability to handle complex, non-linear relationships within MFCC-based features, making them ideal for our goal of accurately classifying insect species. Additionally, Decision Trees are computationally efficient, providing a straightforward yet powerful model for this task.

To ensure robust model evaluation, we employed a cross-validation approach by dividing the dataset into five unique training and testing conditions. In each condition, one audio clip was reserved as the test set, while the remaining clips were used for training. This strategy allowed us to thoroughly assess the model’s performance across varied subsets, offering a comprehensive view of its consistency in accuracy across different data splits. In addition to evaluating the model's performance across different training-test splits, we also investigated how the number of MFCC features impacts the accuracy of the Decision Tree model. To achieve this, we performed experiments using 10, 20, 30, and 40 MFCC features.

For each condition, we trained the Decision Tree model using the \texttt{DecisionTreeClassifier} from the \texttt{scikit-learn} library. By leveraging the mean MFCC feature vectors from the training data, the model was able to learn decision rules tailored to distinguishing between the insect classes. The \texttt{fit()} method facilitated the training, enabling the model to capture unique patterns in the MFCC data that distinguish each insect sound class effectively.

After training, we evaluated the model by generating predictions on the held-out test clip for each condition. Predictions were made using the \texttt{predict()} method, and we calculated accuracy scores by comparing these predictions to the true labels of the test data. The \texttt{accuracy\_score()} function provided a direct measure of the model’s accuracy, and scores were recorded for each condition to enable a thorough analysis of performance.

\begin{figure}[ht]
    \centering
    \includegraphics[width=0.45\textwidth]{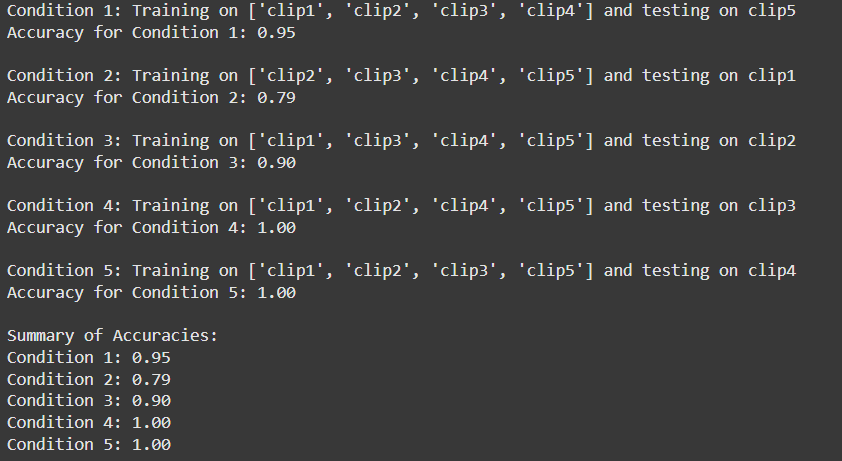}
    \caption{Result of Decision Tree with different conditions}
    \label{fig:dtu10}
\end{figure}

The accuracy results for each condition when extracting 10 features using mfcc were as follows (See Figure \ref{fig:dtu10}):

\begin{itemize}
    \item \textit{Condition 1}: Training on clips 1-4 and testing on clip 5, with an accuracy of 95\%.
    \item \textit{Condition 2}: Training on clips 2-4 and clip 5, testing on clip 1, with an accuracy of 79\%.
    \item \textit{Condition 3}: Training on clips 1, 3-5, testing on clip 2, with an accuracy of 90\%.
    \item \textit{Condition 4}: Training on clips 1-2, 4-5, testing on clip 3, with an accuracy of 100\%.
    \item \textit{Condition 5}: Training on clips 1-3, testing on clip 4, with an accuracy of 100\%.
\end{itemize}

\begin{figure}[ht]
    \centering
    \includegraphics[width=0.45\textwidth]{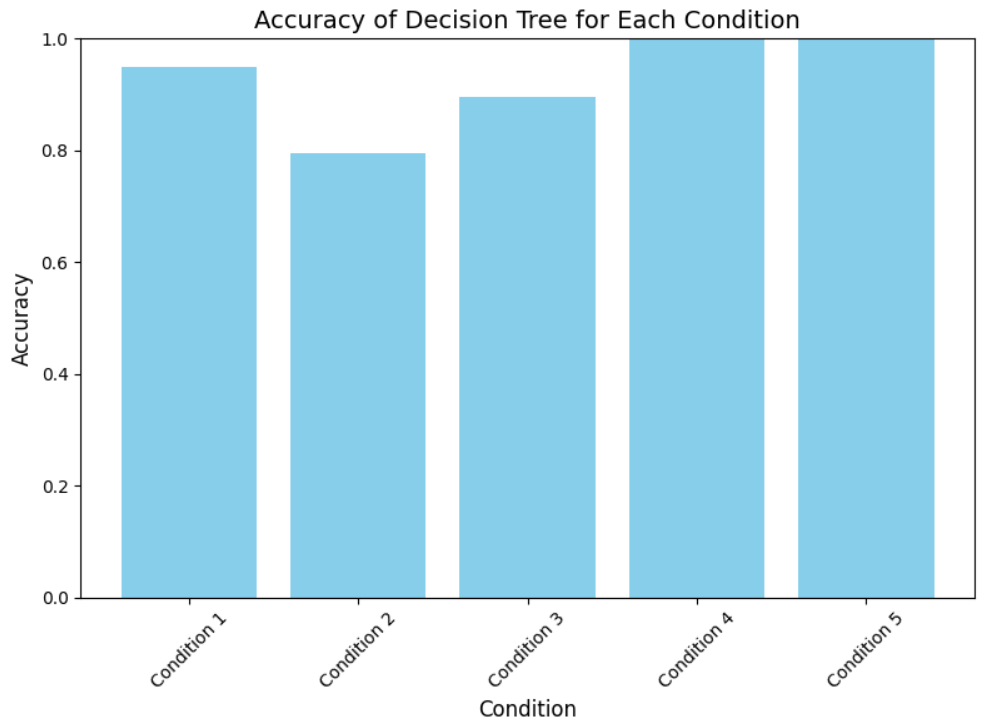}
    \caption{Accuracy of Decision Tree for each condition}
    \label{fig:dtc10}
\end{figure}

The Figure \ref{fig:dtc10} shows the boxplot for accuracy of the Decision Tree model for each condition.

\begin{figure}[ht]
    \centering
    \includegraphics[width=0.45\textwidth]{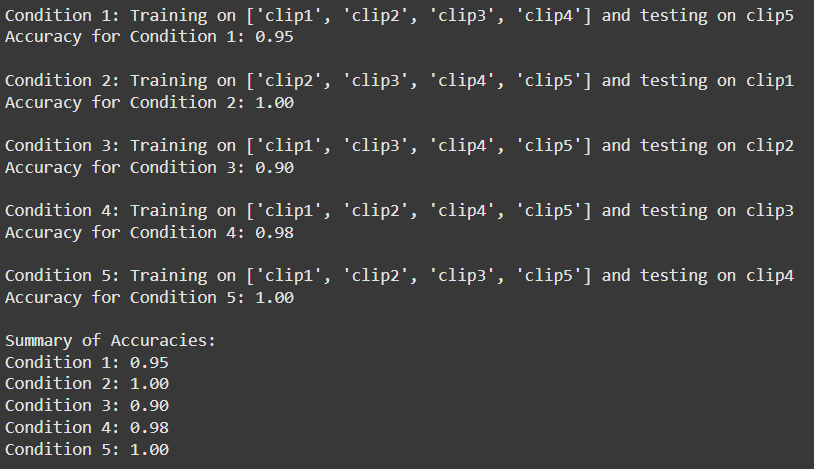}
    \caption{Result of Decision Tree with different conditions}
    \label{fig:dtu}
\end{figure}

The accuracy results for each condition when extracting 20 features using mfcc were as follows (See Figure \ref{fig:dtu}):

\begin{itemize}
    \item \textit{Condition 1}: Training on clips 1-4 and testing on clip 5, with an accuracy of 95\%.
    \item \textit{Condition 2}: Training on clips 2-4 and clip 5, testing on clip 1, with an accuracy of 100\%.
    \item \textit{Condition 3}: Training on clips 1, 3-5, testing on clip 2, with an accuracy of 90\%.
    \item \textit{Condition 4}: Training on clips 1-2, 4-5, testing on clip 3, with an accuracy of 98\%.
    \item \textit{Condition 5}: Training on clips 1-3, testing on clip 4, with an accuracy of 100\%.
\end{itemize}

\begin{figure}[ht]
    \centering
    \includegraphics[width=0.45\textwidth]{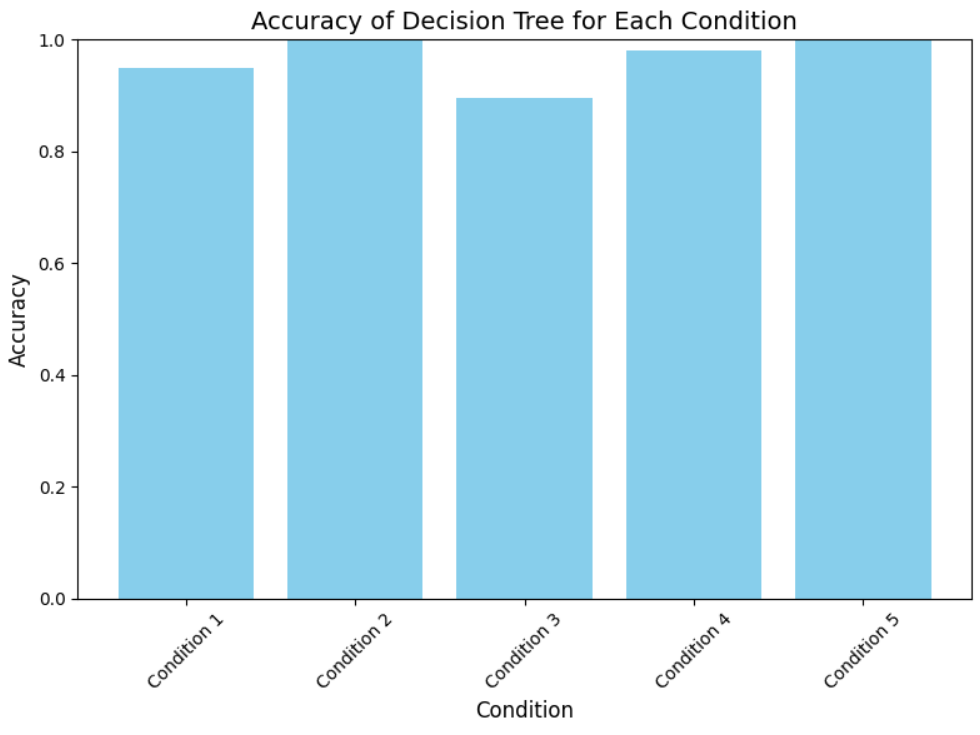}
    \caption{Accuracy of Decision Tree for each condition}
    \label{fig:dtc}
\end{figure}

The Figure \ref{fig:dtc} shows the boxplot for accuracy of the Decision Tree model for each condition. 

\begin{figure}[ht]
    \centering
    \includegraphics[width=0.45\textwidth]{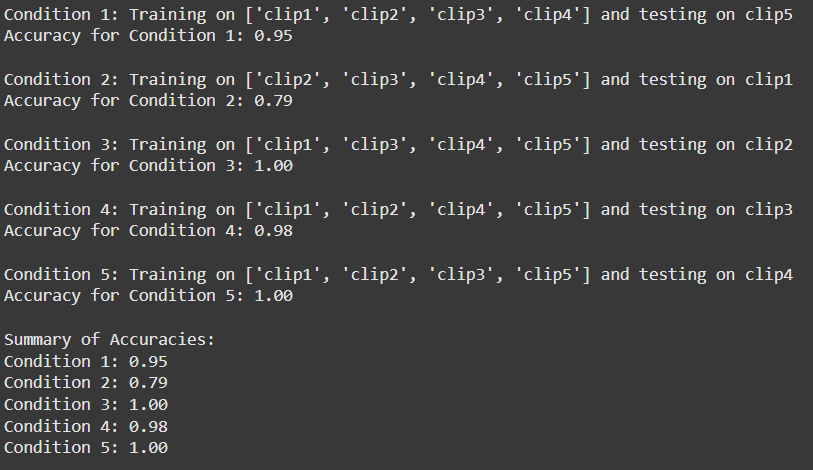}
    \caption{Result of Decision Tree with different conditions}
    \label{fig:dtu30}
\end{figure}

The accuracy results for each condition when extracting 30 features using mfcc were as follows (See Figure \ref{fig:dtu30}):

\begin{itemize}
    \item \textit{Condition 1}: Training on clips 1-4 and testing on clip 5, with an accuracy of 95\%.
    \item \textit{Condition 2}: Training on clips 2-4 and clip 5, testing on clip 1, with an accuracy of 79\%.
    \item \textit{Condition 3}: Training on clips 1, 3-5, testing on clip 2, with an accuracy of 100\%.
    \item \textit{Condition 4}: Training on clips 1-2, 4-5, testing on clip 3, with an accuracy of 98\%.
    \item \textit{Condition 5}: Training on clips 1-3, testing on clip 4, with an accuracy of 100\%.
\end{itemize}

\begin{figure}[ht]
    \centering
    \includegraphics[width=0.45\textwidth]{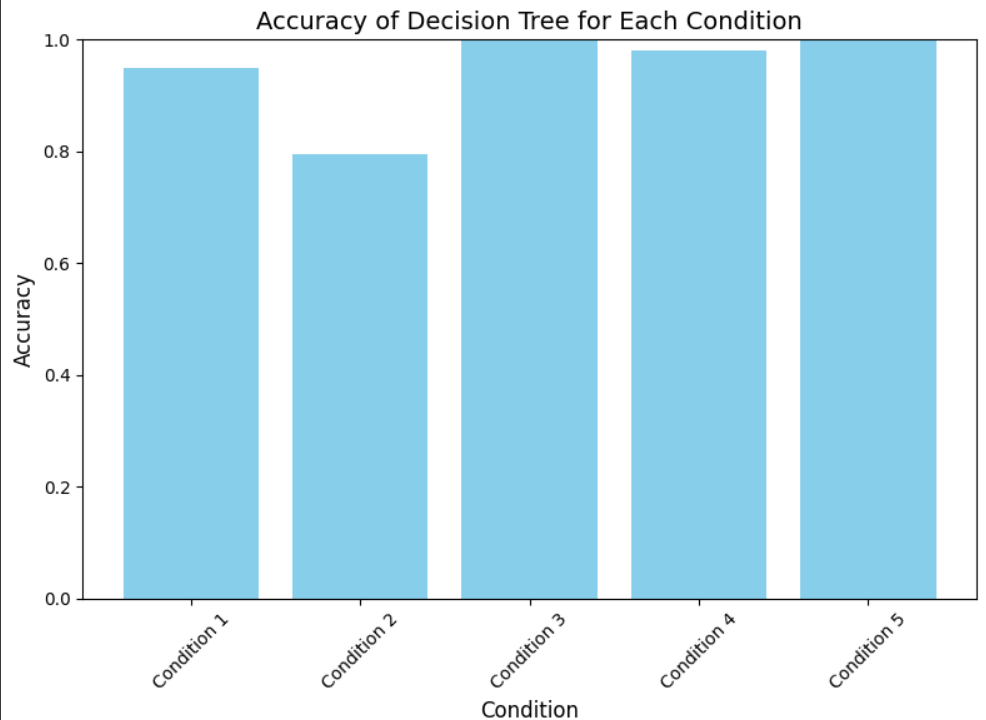}
    \caption{Accuracy of Decision Tree for each condition}
    \label{fig:dtcomp30}
\end{figure}

The Figure \ref{fig:dtcomp30} shows the boxplot for accuracy of the Decision Tree model for each condition. 

\begin{figure}[ht]
    \centering
    \includegraphics[width=0.45\textwidth]{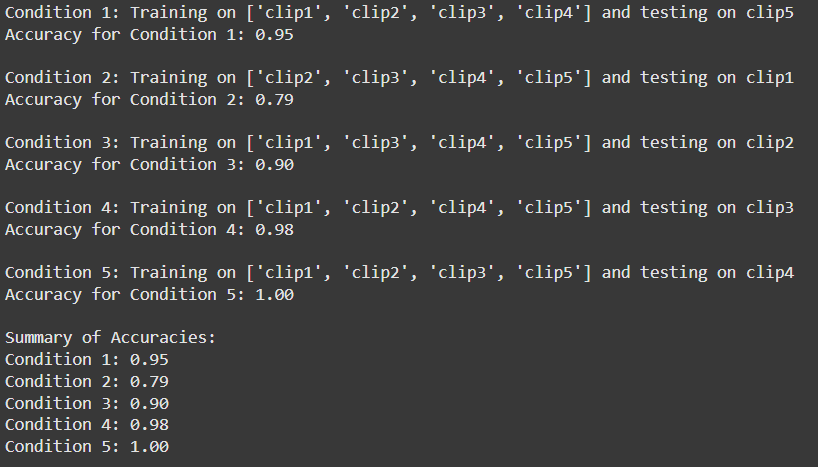}
    \caption{Result of Decision Tree with different conditions}
    \label{fig:dtu40}
\end{figure}

The accuracy results for each condition when extracting 40 features using mfcc were as follows (See Figure \ref{fig:dtu40}):

\begin{itemize}
    \item \textit{Condition 1}: Training on clips 1-4 and testing on clip 5, with an accuracy of 95\%.
    \item \textit{Condition 2}: Training on clips 2-4 and clip 5, testing on clip 1, with an accuracy of 79\%.
    \item \textit{Condition 3}: Training on clips 1, 3-5, testing on clip 2, with an accuracy of 90\%.
    \item \textit{Condition 4}: Training on clips 1-2, 4-5, testing on clip 3, with an accuracy of 98\%.
    \item \textit{Condition 5}: Training on clips 1-3, testing on clip 4, with an accuracy of 100\%.
\end{itemize}

\begin{figure}[ht]
    \centering
    \includegraphics[width=0.45\textwidth]{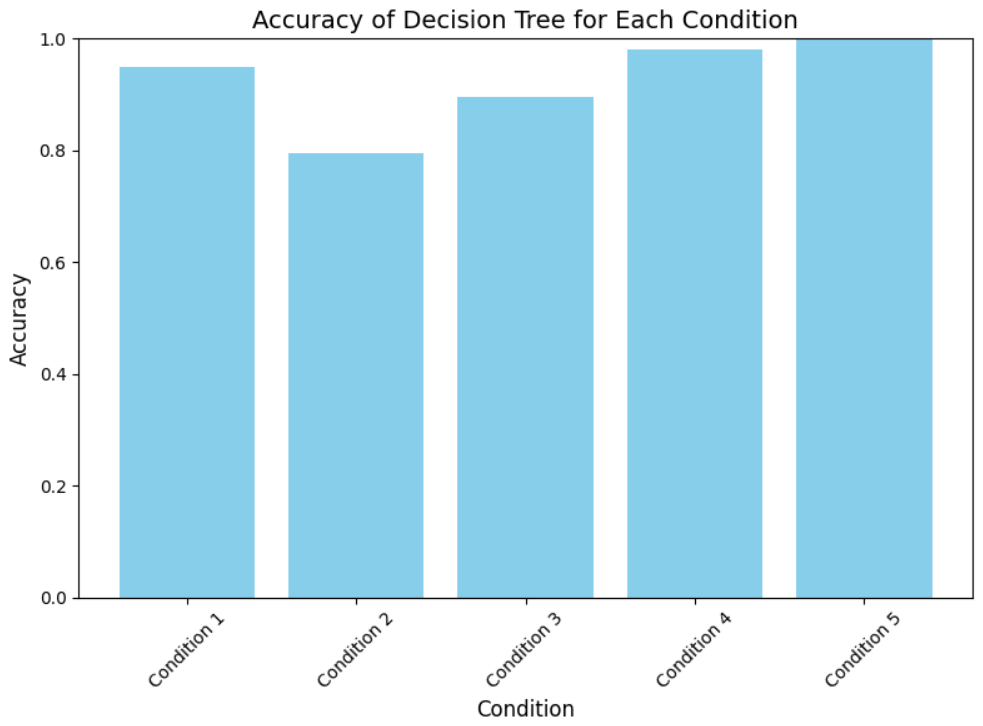}
    \caption{Accuracy of Decision Tree for each condition}
    \label{fig:dtc40}
\end{figure}

The Figure \ref{fig:dtc40} shows the boxplot for accuracy of the Decision Tree model for each condition.

\begin{figure}
    \centering
    \includegraphics[width=0.45\textwidth]{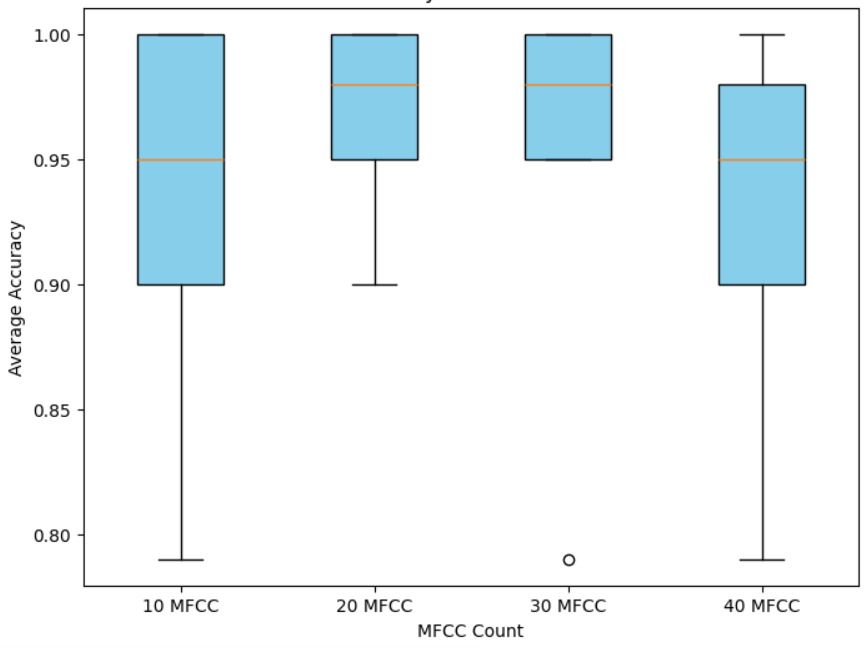}
    \caption{Accuracy Vs MFCC count}
    \label{fig:comparison}
\end{figure}

 The boxplot in Figure \ref{fig:comparison} visualizes the distribution of accuracy scores across these different MFCC feature sets, offering a comparison of how the model's performance varies with the number of features used for training. The plot highlights the trends and variability in accuracy for each MFCC count, providing insights into which configuration yields the best balance between accuracy and computational efficiency.

\begin{figure}
    \centering
    \includegraphics[width=0.45\textwidth]{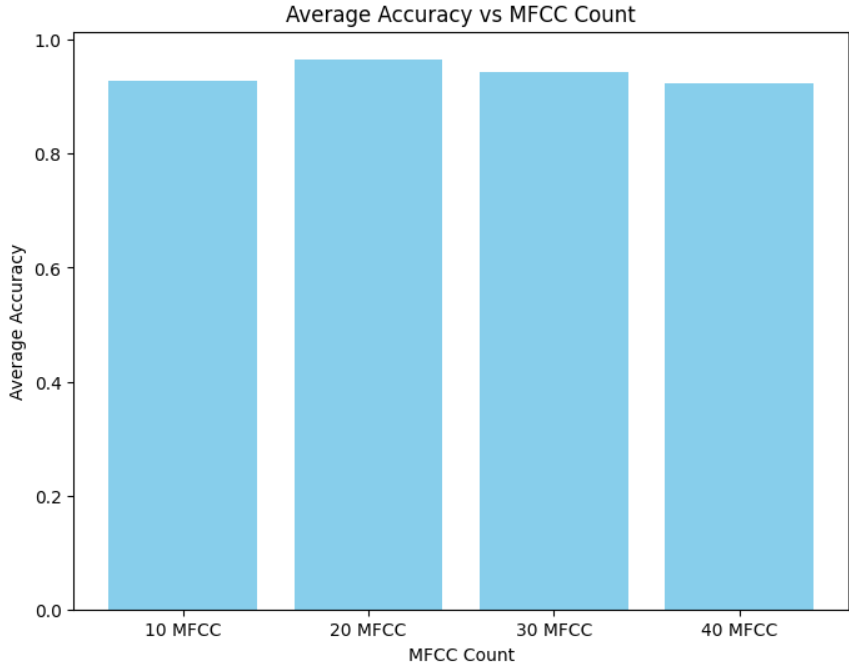}
    \caption{ Average accuracy Vs MFCC count }
    \label{fig:compbar}
\end{figure}

 The bar graph in Figure \ref{fig:compbar} presents the average accuracy achieved by the model for different MFCC feature counts (10, 20, 30, and 40). By consolidating the performance metrics into a single average value for each configuration, the graph offers a clear and straightforward comparison of the model's accuracy across varying MFCC dimensions. This visualization helps identify the optimal number of MFCC features for balancing model performance and feature complexity, emphasizing the trend in average accuracy as the feature count increases.

\section{Distribution of Instances Across Insect Classes}

\begin{table}
    \centering
    \refstepcounter{table}
    \textbf{\large Table I: Instances from Each Clip Across Insect Classes} \\[1ex]
    \begin{tabular}{|c|c|c|c|c|}
        \hline
        \textbf{Clip} & \textbf{Barkbeetle} & \textbf{Cicada} & \textbf{Cricket} & \textbf{Termite} \\ [0.5ex]
        \hline
        Clip 1 & 21 & 8 & 4 & 6 \\ \hline
        Clip 2 & 16 & 3 & 5 & 5 \\ \hline
        Clip 3 & 21 & 9 & 7 & 15 \\ \hline
        Clip 4 & 18 & 9 & 7 & 6 \\ \hline
        Clip 5 & 21 & 7 & 7 & 5 \\ \hline
    \end{tabular}
    \caption*{Table showing the number of instances for each original clip across the four insect classes.}
    \label{table:instances}
\end{table}

This section presents the distribution of instances extracted from audio clips across four insect classes: Barkbeetle, Cicada, Cricket, and Termite. The data is organized into five clips (Clip 1 to Clip 5), with each clip containing a varying number of instances for the respective classes. Barkbeetle consistently has the highest number of instances across all clips, while the distribution of instances for the other classes varies significantly. For example, Clip 3 exhibits the highest number of Termite instances, whereas Cicada instances are comparatively fewer across all clips. This distribution provides valuable insight into the representation of each class within the dataset, which is crucial for training and evaluating classification models effectively. The detailed breakdown is summarized in Table \ref{table:instances}, aiding in understanding the data balance and its potential impact on the classification task.

\section{Visualization of Insect Audio Clips Using UMAP}

In this section, we focus on extracting meaningful features from audio recordings of insect species and visualizing their distributions using UMAP (Uniform Manifold Approximation and Projection). The dataset consists of audio clips categorized into four insect classes: Barkbeetle, Cicada, Cricket, and Termite. Each class contains segmented audio files organized into folders labeled by clip\#. To capture the spectral and temporal characteristics of these recordings, we extract 40 Mel-Frequency Cepstral Coefficients (MFCCs) from each audio file. These MFCC features are then reduced to a 2-dimensional space using UMAP for visualization. Separate UMAP scatter plots are generated for each class, with points color-coded according to their respective clip\# labels, enabling us to explore the feature clustering and relationships among the audio samples within each insect species. The UMAP visualizations for the Barkbeetle, Cicada, Cricket, and Termite classes reveal distinct groupings within each insect class, showcasing the ability of the MFCC features to capture unique spectral and temporal characteristics. The plots also illustrate overlaps and separations between clips, offering insights into intra-class variability and potential challenges in classification. This analysis helps in understanding the feature space and the relationships among audio samples within each class. The UMAP scatter plots for each class are shown in Figures \ref{fig:umpabb}, \ref{fig:umapci}, \ref{fig:umapcr}, \ref{fig:umapt} respectively.

\section{Visualization of Insect Audio Clips Using UMAP}

In this section, we focus on extracting meaningful features from audio recordings of insect species and visualizing their distributions using UMAP (Uniform Manifold Approximation and Projection). The dataset consists of audio clips categorized into four insect classes: Barkbeetle, Cicada, Cricket, and Termite. Each class contains segmented audio files organized into folders labeled by clip\#. To capture the spectral and temporal characteristics of these recordings, we extract 40 Mel-Frequency Cepstral Coefficients (MFCCs) from each audio file. These MFCC features are then reduced to a 2-dimensional space using UMAP for visualization.

\begin{figure}
    \centering
    \includegraphics[width=0.45\textwidth]{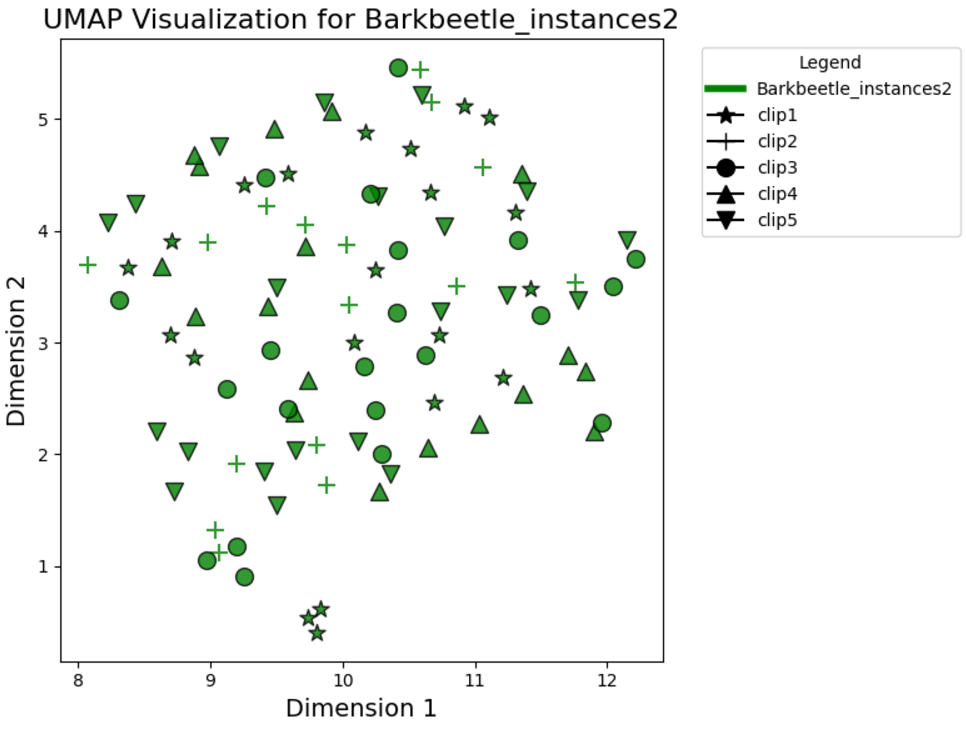}
    \caption{UMAP Visualization for Barkbeetle}
    \label{fig:umpabb}
\end{figure}

\begin{figure}
    \centering
    \includegraphics[width=0.45\textwidth]{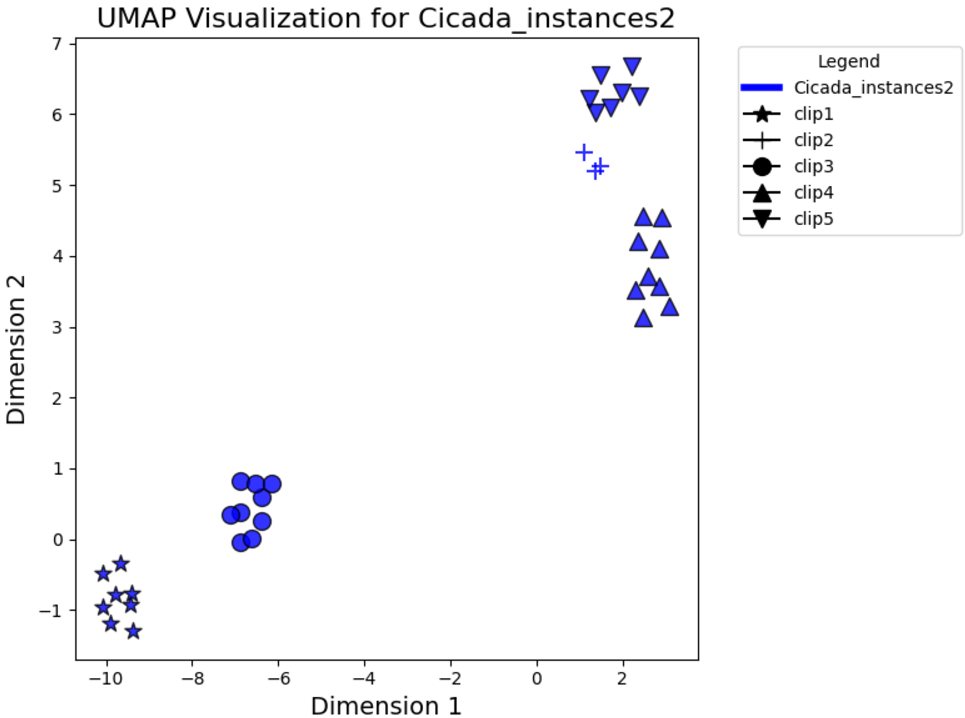}
    \caption{UMAP Visualization for Cicada}
    \label{fig:umapci}
\end{figure}

\begin{figure}
    \centering
    \includegraphics[width=0.45\textwidth]{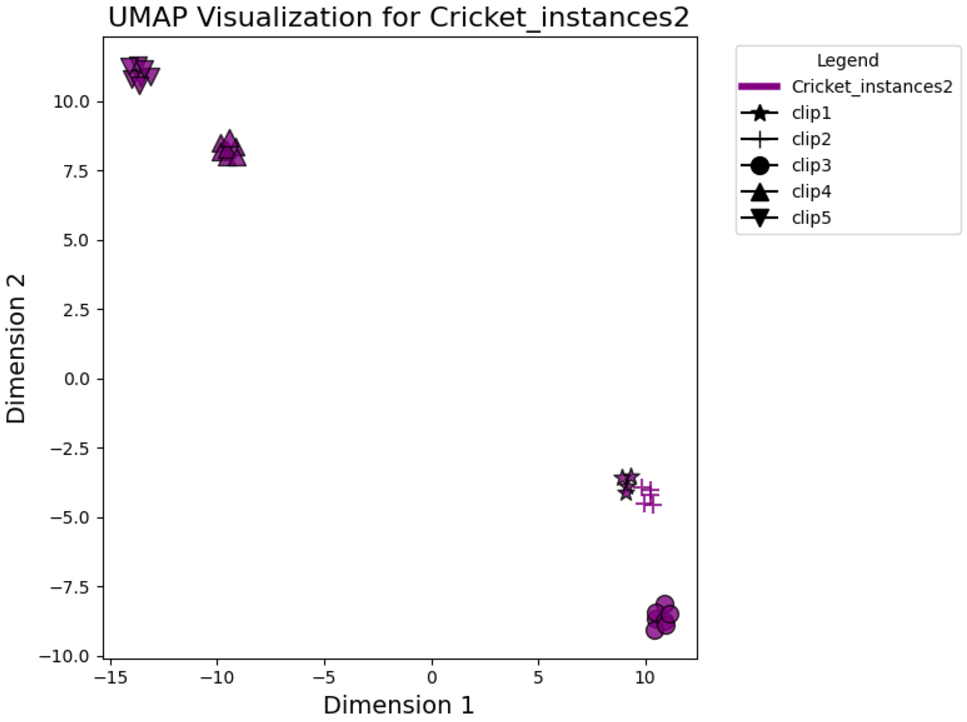}
    \caption{UMAP Visualization for Cricket}
    \label{fig:umapcr}
\end{figure}

\begin{figure}
    \centering
    \includegraphics[width=0.45\textwidth]{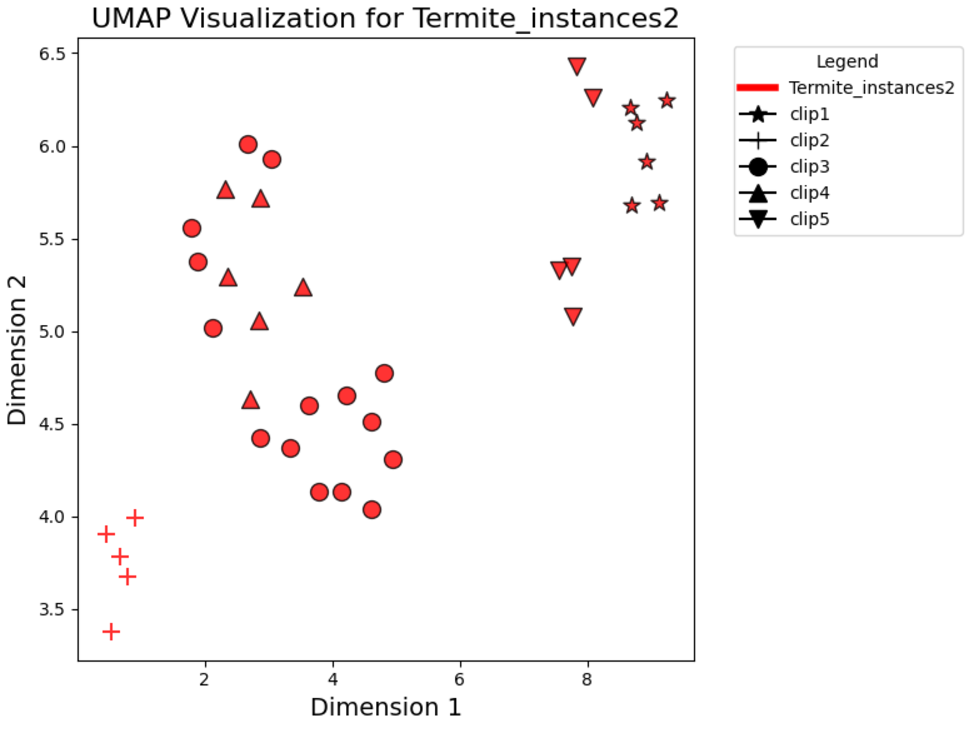}
    \caption{UMAP Visualization for Termite}
    \label{fig:umapt}
\end{figure}

Separate UMAP scatter plots are generated for each class, with points color-coded according to their respective clip\# labels and styled with distinct markers. This approach enables us to explore the feature clustering and relationships among the audio samples within each insect species. The UMAP visualizations for the Barkbeetle, Cicada, Cricket, and Termite classes reveal distinct groupings within each insect class, showcasing the ability of the MFCC features to capture unique spectral and temporal characteristics. The plots also illustrate overlaps and separations between clips, offering insights into intra-class variability and potential challenges in classification. The UMAP scatter plots for each class are shown in Figures \ref{fig:umpabb}, \ref{fig:umapci}, \ref{fig:umapcr}, \ref{fig:umapt} respectively.

\begin{figure}
    \centering
    \includegraphics[width=0.45\textwidth]{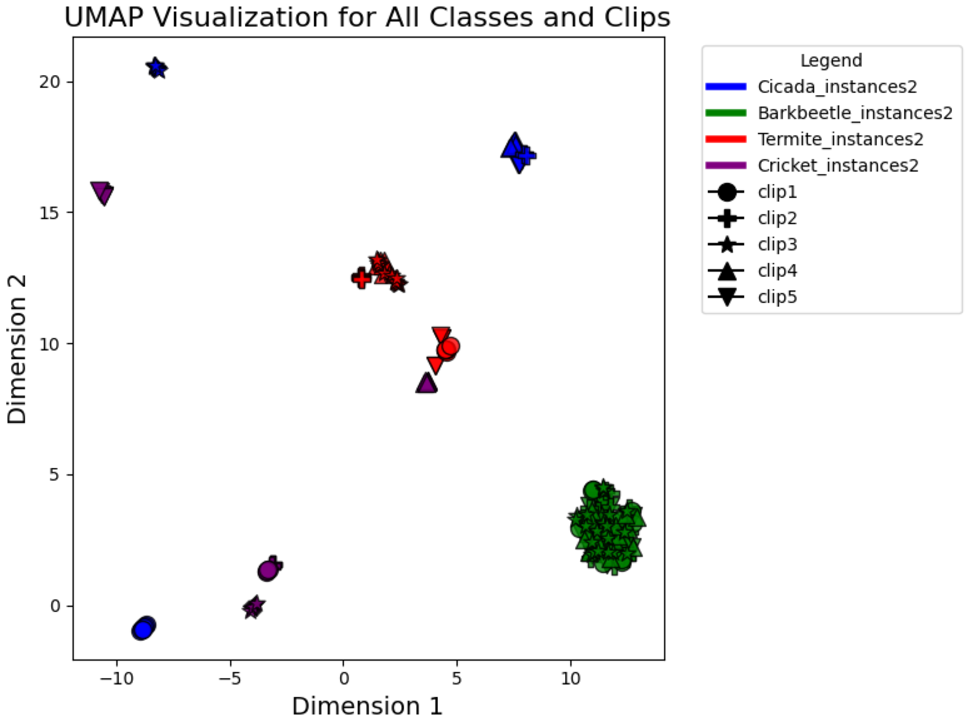}
    \caption{Combined UMAP Visualization for All Classes and Clips}
    \label{fig:allw}
\end{figure}

In addition to the individual visualizations for each insect class, a combined UMAP plot was created to analyze the inter-class relationships and overall feature distribution across all insect species and clips. This comprehensive visualization integrates all the audio samples into a single plot, with each insect class distinguished by a unique color and each clip by a specific marker style. The combined UMAP plot, shown in Figure \ref{fig:allw}, highlights the separations and overlaps between the classes, providing a holistic perspective on how the features cluster in the reduced-dimensional space. This visualization emphasizes the distinctiveness of the classes while also revealing areas of potential inter-class confusion, thereby guiding further refinements in feature extraction and classification strategies.


\section{Feature Importance Analysis Using Random Forests}

\begin{figure}
    \centering
    \includegraphics[width=0.5\textwidth]{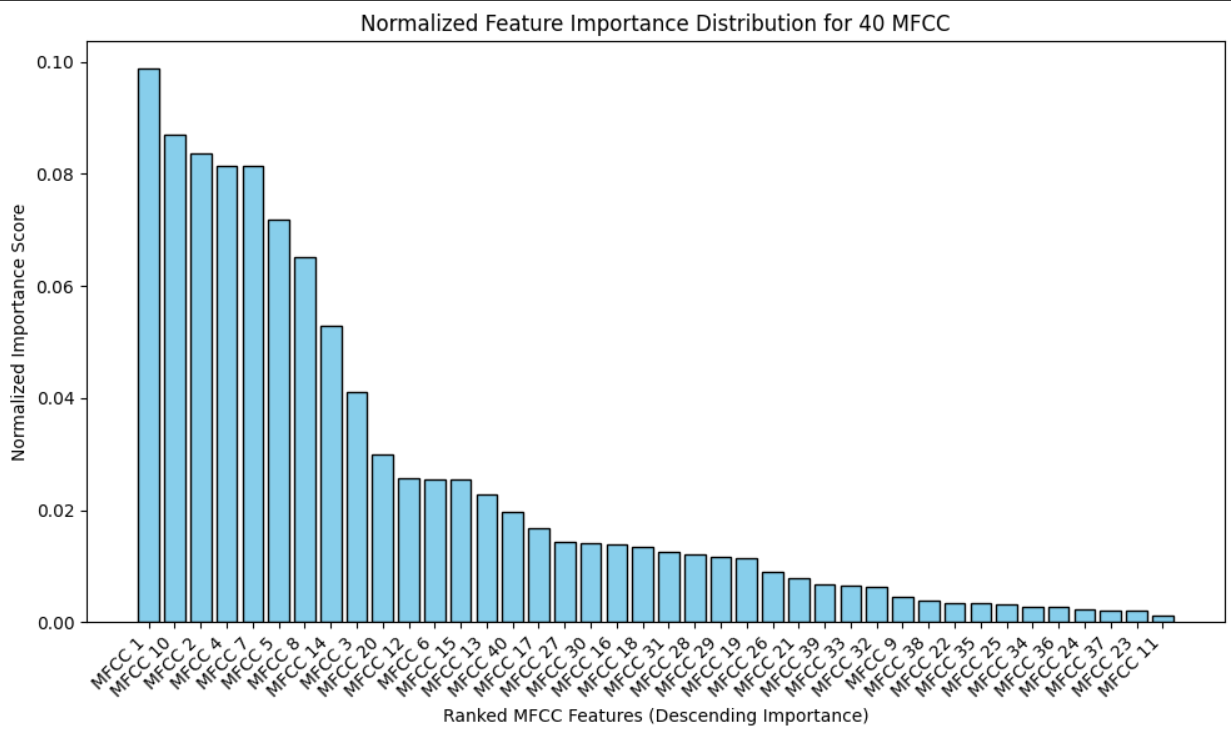}
    \caption{Normalized feature importance distribution for 40 MFCC}
    \label{fig:40mfcc}
\end{figure}

In this section, we analyzed the importance of MFCC features using a Random Forest classifier to evaluate their contribution to classification tasks as shown in Figure ~\ref{fig:40mfcc}. Initially, we focused on computing and visualizing the normalized feature importance of 40 MFCCs. The importance scores were normalized to a scale where their total summed to 1, allowing for a direct comparison of each feature's relative significance. By sorting the features in descending order of importance and plotting a bar graph, we identified the most influential MFCC features. This visualization highlights the specific features that play a critical role in the model's predictions, aiding in feature selection and prioritization for further analysis.

\begin{figure}
    \centering
    \includegraphics[width=0.45\textwidth]{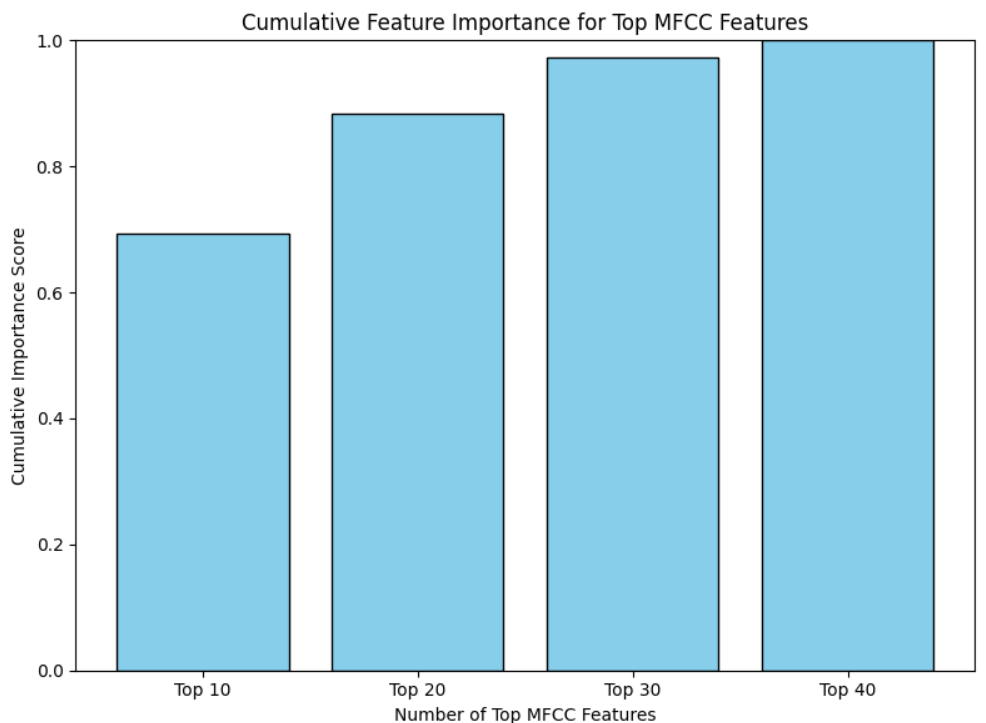}
    \caption{Cumulative feature importance for top MFCC features}
    \label{fig:topmfcc}
\end{figure}

Next, we examined the cumulative contribution of subsets of MFCC features, focusing on the top 10, 20, 30, and 40 features as shown in Figure ~\ref{fig:topmfcc}. By summing the importance scores for these groups, we assessed how the contribution scales as additional features are included. A bar graph was used to illustrate these cumulative scores, providing a clear visual representation of the incremental value added by larger sets of MFCC features. This approach helps determine the optimal number of features needed for effective classification, balancing model performance and computational efficiency.

\section{Multiple Model Evaluation and Comparison Using Original Audio}

In this study, we evaluated the performance of multiple machine learning models for audio-based classification of insect species using original audio data. The dataset consisted of audio clips organized into four insect classes: Barkbeetle, Cicada, Cricket, and Termite. A leave-one-clip-out cross-validation strategy was employed, wherein five conditions were designed by reserving one clip for testing while using the remaining four clips for training. Audio features were extracted using 40-dimensional Mel-Frequency Cepstral Coefficients (MFCCs), which served as inputs to the models. 

\begin{figure}
    \centering
    \includegraphics[width=0.45\textwidth]{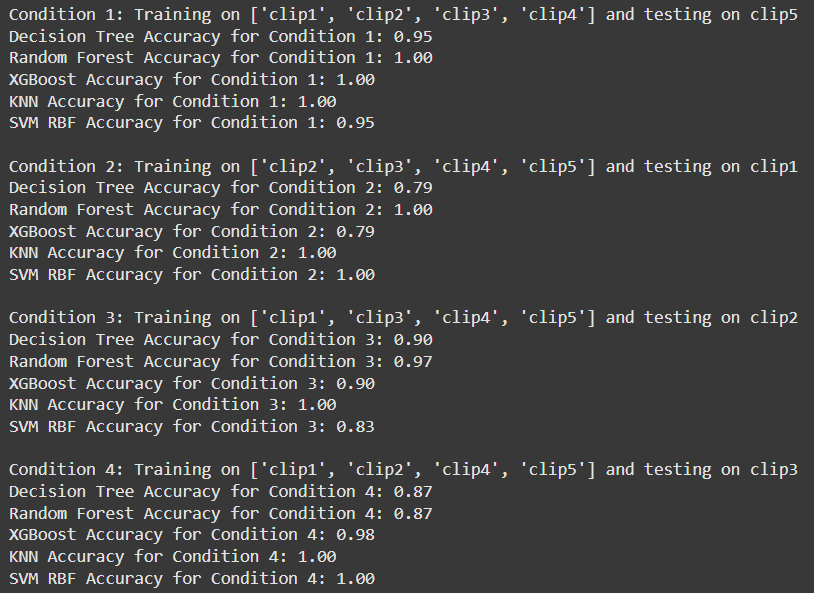}
    \label{fig:ogdata}
\end{figure}
\begin{figure}
    \centering
    \includegraphics[width=0.45\textwidth]{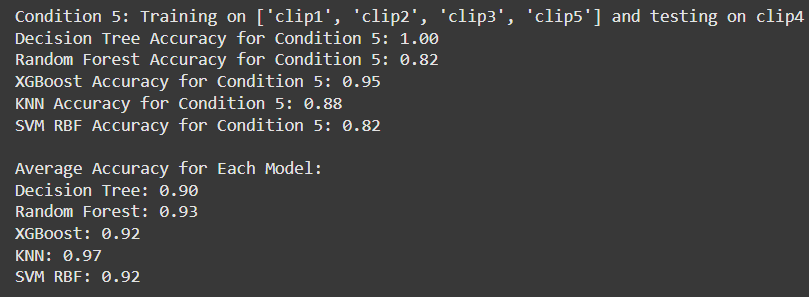}
    \caption{Accuracies for different models}
    \label{fig:ogdata1}
\end{figure}

The classification models evaluated included Decision Tree, Random Forest, XGBoost, K-Nearest Neighbors (KNN), and Support Vector Machine (SVM) with an RBF kernel. To ensure consistent labeling, class labels were encoded numerically using a LabelEncoder. The models were trained and tested under each condition, and their performance was quantified using accuracy scores. The results demonstrated that all models achieved high accuracy across the various conditions, with notable variability in performance between them. KNN consistently achieved the highest average accuracy of 0.97 across all conditions, indicating its robustness in capturing patterns within the MFCC features. Random Forest and XGBoost models followed closely, with average accuracies of 0.93 and 0.92, respectively. SVM with an RBF kernel also achieved an average accuracy of 0.92, while Decision Tree recorded a slightly lower average accuracy of 0.90 (See Figure \ref{fig:ogdata1}).

The evaluation highlights the strong predictive capabilities of all the models employed, with KNN emerging as the most reliable for this task. The findings demonstrate the potential of these machine learning approaches in leveraging MFCC features for the classification of insect audio recordings and provide a comparative understanding of their strengths under varying data splits.

\section{Audio Data Augmentation and Multiple Model Evaluation \& Comparison}
In this section, we implemented audio data augmentation to improve the diversity of our insect sound dataset and subsequently evaluated several machine learning models for classification tasks. The dataset consists of audio recordings from four classes of insects: Barkbeetle, Cicada, Cricket, and Termite. To augment the data, we applied two primary techniques: time stretching and pitch shifting. Time stretching was achieved by varying the playback speed of the audio, and pitch shifting was done by modifying the frame rate of the audio, both using the pydub library. These augmentations aimed to simulate different environmental conditions and variations in insect sounds, thereby enriching the dataset for model training.

We organized the augmented data into separate folders for each insect class, ensuring that the augmented instances were stored alongside the original recordings. The augmented audio files were then processed to extract Mel-frequency cepstral coefficients (MFCCs), a commonly used feature for audio classification tasks. MFCCs were extracted using the librosa library, and the mean values of 40 MFCC coefficients were used as the feature set for each audio clip. These features served as the input for our machine learning models.

To evaluate the effectiveness of the augmentation and to benchmark different classification models, we performed a series of experiments using multiple train-test splits. The train-test splits were designed to test the models' ability to generalize by using different combinations of clips for training and testing. We implemented five distinct models: Decision Tree, Random Forest, XGBoost, K-Nearest Neighbors (KNN), and Support Vector Machine (SVM) with an RBF kernel. For each condition, we trained the models on the training set and evaluated them on the corresponding test set, using accuracy as the evaluation metric.

\begin{figure}
    \centering
    \includegraphics[width=0.45\textwidth]{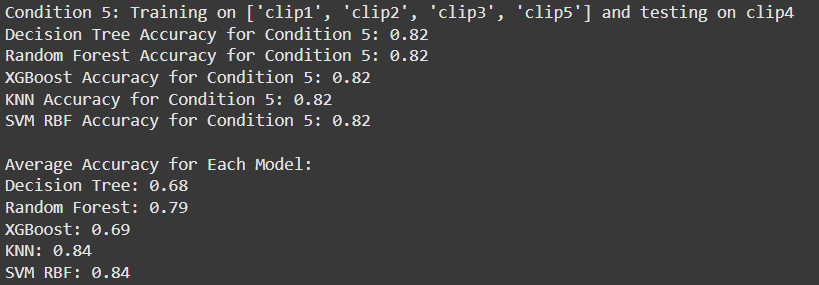}
    \caption{Accuracies for different models}
    \label{fig:c2}
\end{figure}

The results showed varied performance across the different models and conditions. For Condition 1, where the model was trained on clips 1 to 4 and tested on clip 5, KNN and SVM RBF performed the best with accuracy scores of 0.82, while Decision Tree, Random Forest, and XGBoost performed moderately with accuracies of 0.55, 0.65, and 0.65, respectively. In Condition 2, training on clips 2 to 5 and testing on clip 1, Random Forest, KNN, and SVM RBF performed well with accuracies of 0.85, whereas XGBoost showed a lower accuracy of 0.38. Condition 3, testing on clip 2 while training on the other clips, saw Decision Tree outperforming with 0.83 accuracy, while other models showed lower but still respectable results. Condition 4, training on clips 1, 2, 4, and 5 and testing on clip 3, highlighted KNN and SVM RBF as the top performers, both achieving 1.00 accuracy. Finally, in Condition 5, where clip 4 was used as the test set, KNN, Random Forest, and SVM RBF showed consistent performance with accuracies of 0.82 (See Figure \ref{fig:c2}).
\begin{figure}
    \centering
    \includegraphics[width=0.45\textwidth]{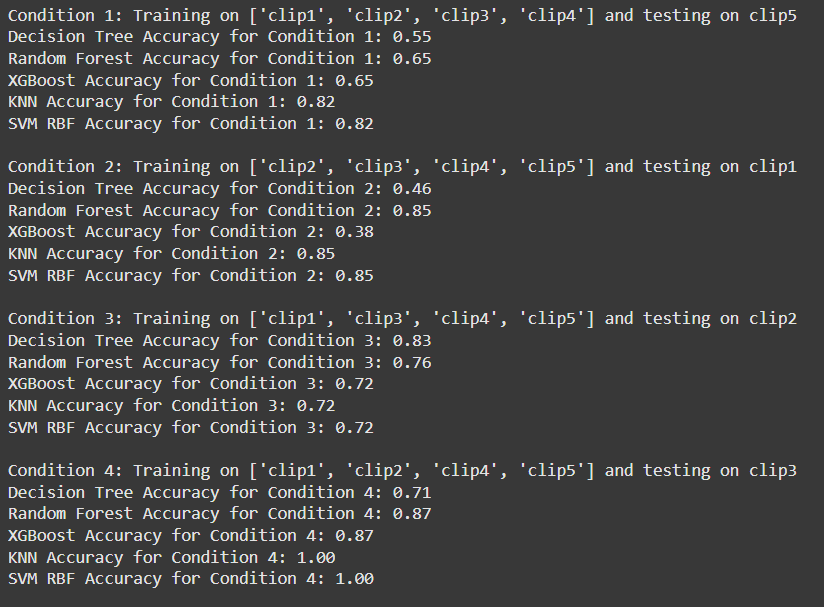}
    \label{fig:C1}
\end{figure}

The average accuracy results across all conditions revealed that KNN and SVM RBF performed the best, both achieving an average accuracy of 0.84. Random Forest followed with an average of 0.79, while Decision Tree and XGBoost averaged at 0.68 and 0.69, respectively. These results demonstrate that KNN and SVM RBF are more robust to the variations in the dataset introduced by the different training and testing splits, while Random Forest also performed relatively well. The performance differences across models further highlight the importance of model selection in audio classification tasks, as certain models like KNN and SVM RBF may better capture the patterns within the insect sound data.

\section{Analysis of Reduced Model Accuracy Post-Augmentation}

\begin{figure}
    \centering
    \includegraphics[width=0.45\textwidth]{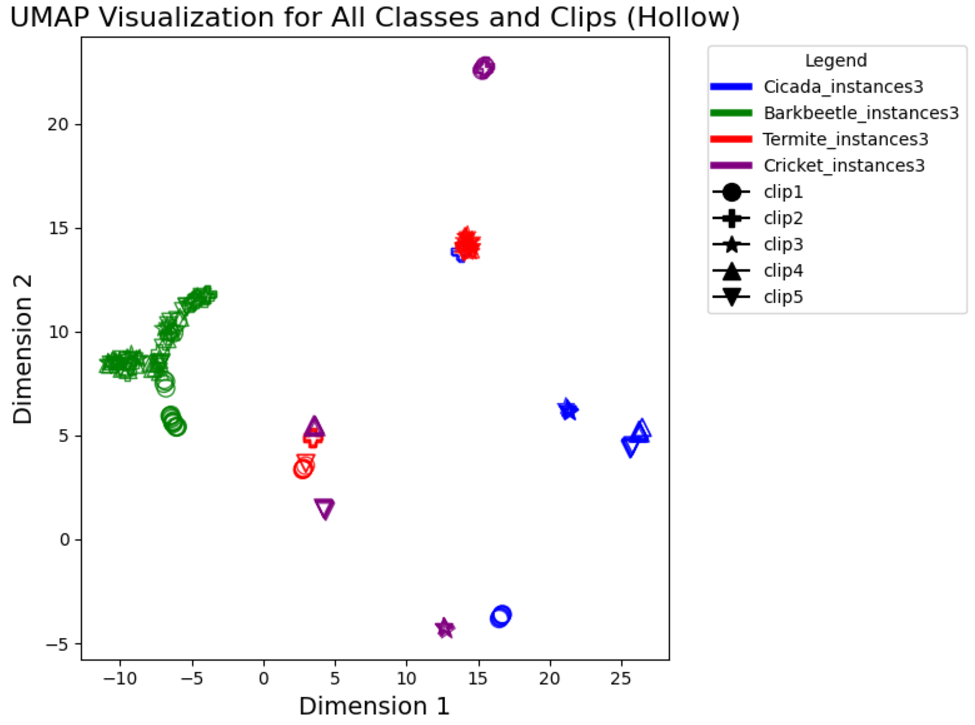}
    \caption{UMAP Visualization after data augmentation}
    \label{fig:aug}
\end{figure}

The observed reduction in accuracy for all models after data augmentation can be attributed to several factors, as highlighted by the UMAP visualization in Figure~\ref{fig:aug}. Augmentation techniques, such as time stretching and pitch shifting, introduced variability into the dataset to simulate diverse environmental conditions and insect sound patterns. However, this variability likely introduced overlaps in feature space across classes, complicating the classification task for machine learning models.

From Figure~\ref{fig:aug}, it is evident that instances of certain classes, such as "Cicada\_instances3" and "Termite\_instances3", show significant overlap in the two-dimensional feature space. This overlap suggests that the augmented audio features for these classes share similar patterns, reducing the models' ability to clearly distinguish between them. Such overlaps may result from the limited distinctiveness of MFCC features when subjected to transformations like pitch shifting, which can cause sounds from different classes to converge in feature space.

Another contributing factor to the reduced accuracy is the potential misalignment between the augmented data and the inherent characteristics of the original dataset. While augmentation enriches the dataset, it can inadvertently create unrealistic patterns or distortions that the models have not been trained to handle effectively. This issue is particularly pronounced for models like Decision Tree and XGBoost, which rely on clear separations in feature space to perform well, as indicated by their lower average accuracies.

Additionally, the increased complexity of the dataset after augmentation may have heightened the models' sensitivity to noise. This is evident from the performance drop of models such as Random Forest and KNN, which, although robust to some variability, struggle when classes are no longer well-separated in feature space. Models like SVM with an RBF kernel, which performed well before augmentation, also saw reduced accuracy due to their reliance on hyperplanes to separate classes—an approach that becomes less effective when class boundaries blur.

In summary, the reduction in accuracy across all models can be attributed to the introduction of overlapping class distributions, the challenges of capturing augmented data's variability, and the models' sensitivity to noise and distorted patterns. The findings underscore the trade-offs involved in data augmentation, emphasizing the need for careful selection of augmentation techniques and feature representations to enhance model robustness without compromising performance.

\section{Discussion / Limitations}
In this study, we explored the classification of insect species based on audio recordings, utilizing Mel-Frequency Cepstral Coefficients (MFCCs) and a variety of machine learning models, including Random Forest, K-Nearest Neighbors (KNN), and Support Vector Machine (SVM), among others. The primary objective was to assess the performance of these models in identifying distinct insect species, including Barkbeetle, Cicada, Cricket, and Termite, and to understand the impact of different feature extraction techniques and data augmentation strategies. The results of this study highlight the potential of using MFCC features in conjunction with machine learning models for insect sound classification, but several limitations and challenges emerged that should be considered for future research.

\subsection{Model Performance and Feature Selection}

While most models performed well in terms of classification accuracy, the KNN model stood out with the highest average accuracy of 0.97 across all conditions. This suggests that KNN is particularly effective in capturing the underlying patterns of the insect audio data, possibly due to its ability to consider the local structure of the feature space. On the other hand, models like Decision Tree and XGBoost demonstrated slightly lower accuracy, which may be attributed to their reliance on decision boundaries that may not align well with the intricacies of the MFCC feature space. Random Forest and SVM (RBF kernel) showed moderate performance, highlighting that feature importance and the model's ability to handle complex data distributions are key factors in achieving robust classification.

In terms of feature selection, the Random Forest feature importance analysis provided valuable insights into which MFCC coefficients were most influential for classification. While these findings help refine the feature set, further research could investigate additional feature extraction techniques, such as spectral contrast or chroma features, to potentially enhance classification performance. Additionally, reducing the dimensionality of the feature space, perhaps through techniques like PCA or feature selection, could help streamline the models and improve computational efficiency without sacrificing accuracy.

\subsection{Impact of Data Augmentation}

Data augmentation proved to be a double-edged sword in this study. While augmenting the dataset with time stretching and pitch shifting enhanced its diversity, it also introduced unintended overlaps in the feature space, particularly after augmentation. This overlap made it more challenging for the models to distinguish between certain insect species, leading to a reduction in classification accuracy. For example, the UMAP visualizations post-augmentation showed significant clustering of instances from different classes, which could have contributed to the increased difficulty in achieving high performance after augmentation.

The reduction in accuracy after augmentation underscores a potential limitation of using simple augmentation techniques like time stretching and pitch shifting. These transformations, while helpful in simulating different acoustic environments, may not always reflect realistic variations in insect sounds. Future studies could explore more sophisticated augmentation techniques, such as adding noise or simulating real-world environmental conditions, to create more robust datasets. Additionally, domain-specific data augmentation strategies that consider the biological characteristics of insect sounds could lead to more meaningful transformations and improve model performance.

\subsection{Generalization and Overfitting}

Although the models demonstrated strong performance, it is important to consider the potential for overfitting. The high accuracies observed in certain conditions, particularly with KNN, could suggest that the model is memorizing specific patterns rather than generalizing well to unseen data. This is a common issue when dealing with small datasets, as was the case in this study. The use of cross-validation helped mitigate this risk, but further investigation into the models' ability to generalize to new, unseen recordings is necessary. It would be beneficial to test the models on larger, more diverse datasets to assess their robustness and ensure that they are not overly sensitive to the particularities of the current dataset.

\subsection{Class Imbalance and Intra-Class Variability}

Another limitation of this study is the potential class imbalance in the dataset. If certain insect classes are underrepresented or if there are significant variations in the number of clips per class, this could impact the overall performance of the models. Additionally, the intra-class variability—differences within the same insect class in terms of sound characteristics—was evident in the UMAP visualizations, where some classes showed more pronounced separations than others. This variability can pose challenges in classification, as models may struggle to identify subtle differences within a class, especially when dealing with noisy or inconsistent data.

Future work could address this issue by ensuring a more balanced dataset or by employing techniques like class weighting or oversampling to mitigate the effects of class imbalance. Furthermore, exploring advanced model architectures such as deep learning models (e.g., CNNs or RNNs) could help improve performance by better capturing complex patterns in the audio data and handling intra-class variability more effectively.

\subsection{Future Directions}

Several areas for improvement and exploration remain for future studies. First, additional feature extraction techniques, such as temporal or rhythm-based features, could be incorporated to better capture the dynamic aspects of insect sounds. Furthermore, exploring other machine learning algorithms, including deep learning models, could offer more robust performance, especially when dealing with more complex and diverse datasets.

Moreover, more sophisticated data augmentation strategies, such as adding environmental noise or varying the background acoustics, could provide a better simulation of real-world conditions and improve model robustness. Finally, incorporating metadata about the insect species or environmental context (e.g., time of day, geographic location) could provide additional features that help improve classification accuracy and generalizability.

\subsection{Conclusion}

In summary, this study provides valuable insights into the potential of using MFCC features and machine learning models for insect sound classification. While the models demonstrated strong performance, the results also highlighted important challenges, including the impact of data augmentation, class imbalance, and intra-class variability. Addressing these limitations and exploring new techniques for feature extraction, model selection, and data augmentation will be essential for improving the accuracy and robustness of insect classification models. This research paves the way for further advancements in bioacoustic monitoring and pest management applications, where accurate and reliable insect sound classification can play a key role in ecological studies and pest control efforts.

\section*{Acknowledgment}
We would like to thank Dr. Sudip Vhaduri for his guidance in this work.

\bibliographystyle{IEEEbib}
\bibliography{Reference}

\end{document}